\documentclass[sigconf]{acmart} 

\AtBeginDocument{%
  \providecommand\BibTeX{{%
    \normalfont B\kern-0.5em{\scshape i\kern-0.25em b}\kern-0.8em\TeX}}}

\acmConference[Preprint]{Preprint}{January 13th}{2020}

\begin{document}

\title{Towards Patient Record Summarization Through Joint Phenotype Learning in HIV Patients}

\author{Gal Levy-Fix}
\affiliation{%
  \institution{Department of Biomedical Informatics \\ Columbia University}
  \streetaddress{622 W. 168th Street}
  \city{New York}
  \state{NY}
}
\author{Jason Zucker}
\affiliation{%
  \institution{Division of Infectious Diseases \\Columbia University Medical Center}
  \streetaddress{622 W. 168th Street}
  \city{New York}
  \country{NY}}

\author{Konstantin Stojanovic}
\affiliation{%
  \institution{Department of Neurology \\Columbia University Medical Center}
  \streetaddress{622 W. 168th Street}
  \city{New York}
  \country{NY}}

\author{No\'emie Elhadad}
\affiliation{%
  \institution{Department of Biomedical Informatics \\Columbia University}
  \streetaddress{622 W. 168th Street}
  \city{New York}
  \country{NY}}

\renewcommand{\shortauthors}{Levy-Fix et al.}

\begin{abstract}
Identifying a patient's key problems over time is a common task for providers at the point care, yet a complex and time-consuming activity given current electric health records. To enable a problem-oriented summarizer to identify a patient's comprehensive list of problems and their salience, we propose an unsupervised phenotyping approach that jointly learns a large number of phenotypes/problems across structured and unstructured data. To identify the appropriate granularity of the learned phenotypes, the model is trained on a target patient population of the same clinic. To enable the content organization of a problem-oriented summarizer, the model identifies phenotype relatedness as well. The model leverages a correlated-mixed membership approach with variational inference applied to heterogenous clinical data. In this paper, we focus our experiments on assessing the learned phenotypes and their relatedness as learned from a specific patient population. We ground our experiments in phenotyping patients from an HIV clinic in a large urban care institution (n=7,523), where patients have voluminous, longitudinal documentation, and where providers would benefit from summaries of these patient's medical histories, whether about their HIV or any comorbidities. We find that the learned phenotypes and their relatedness are clinically valid when assessed qualitatively by clinical experts, and that the model surpasses baseline in inferring phenotype-relatedness when comparing to existing expert-curated condition groupings. 

\end{abstract}

%

\keywords{Patient summarization, Graphical models, EHR, Phenotyping}


\maketitle

\section{Introduction}

Electronic health records (EHR) have improved the availability of patient records, but this has not always translated to increased availability of relevant information to clinicians \cite{christensen_Instant_2008}. This is partly because increased amounts of data in EHRs has made it more difficult for clinicians to review patients' previous medical histories and obtain an overview of the patient record \cite{JENSEN201644}. Increased amounts patient data have also raised concerns regarding clinician information overload \cite{farri_Qualitative_2012}, having effects on care quality\cite{cj_Use_2014}, and patient safety\cite{rj_Cognitive_2011}. 

Patient record summarization has been suggested as a valuable tool to support clinicians in making sense of increasingly large patient records\cite{feblowitz_Summarization_2011}. There are a number of open challenges associated with robust summarization of clinical documentation \cite{pivovarov_Automated_2015}, including content selection---identifying the right summary elements at the right granularity in the input patient record--- and content organization---organizing summary output in a coherent and actionable fashion for the clinicians, all the while preserving data provenance. 
Previous work has shown that problem-oriented summary supports the needs of clinicians \cite{weed_Medical_1968,hirsch_HARVEST_2014,lee_Public_2015, li_Impact_2018}. High-throughput computational phenotyping methods are attractive for identifying a patient's problems in a robust and scalable fashion \cite{ho_Limestone_2014, pivovarov_Learning_2015, joshi_Identifiable_2016}. Considering the characteristics of electronic health record (EHR) data (missingness, heterogeneity, uncertainty), Bayesian generative approaches are attractive to handle them and provide easily interpretable outputs that quantify their uncertainty. 

\begin{figure}[h]
	\centering
	\includegraphics[width=\linewidth]{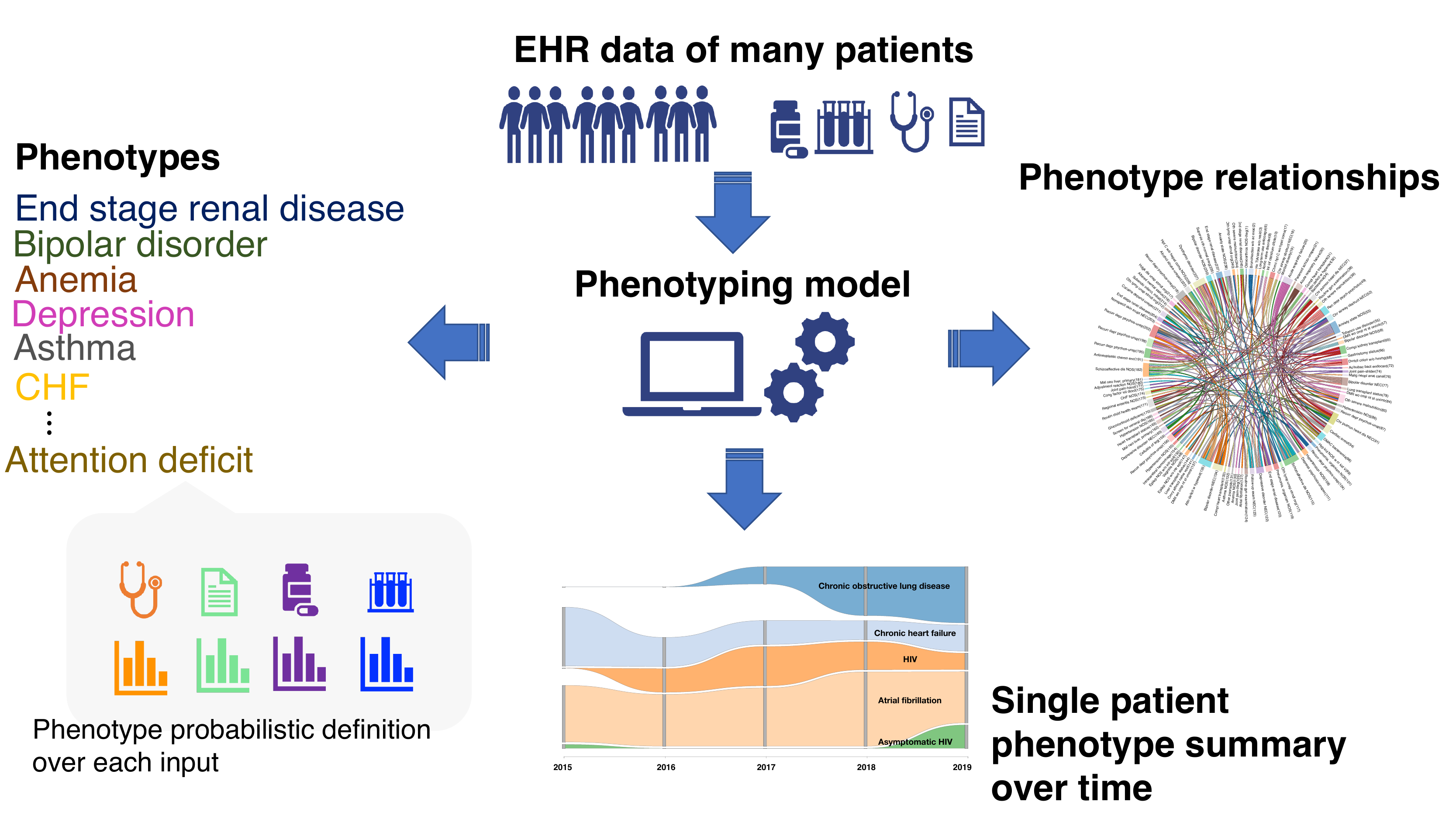}
	\caption{Example of five learned phenotypes and their learned correlations}
	\label{fig:approach}
\end{figure}

To enable a problem-oriented summarizer to identify a target patient's comprehensive list of problems and their salience, we propose a probabilistic machine learning approach that can identify a large number of phenotypes/problems using patients' structured and unstructured data in an unsupervised fashion. The machine learning model is trained on the EHR data of many patients to simultaneously learn probabilistic definitions of many phenotypes at the same time.  Figure~\ref{fig:approach} shows a graphical schema of the proposed approach .

\begin{figure}[h]
	\centering
	\includegraphics[width=\linewidth]{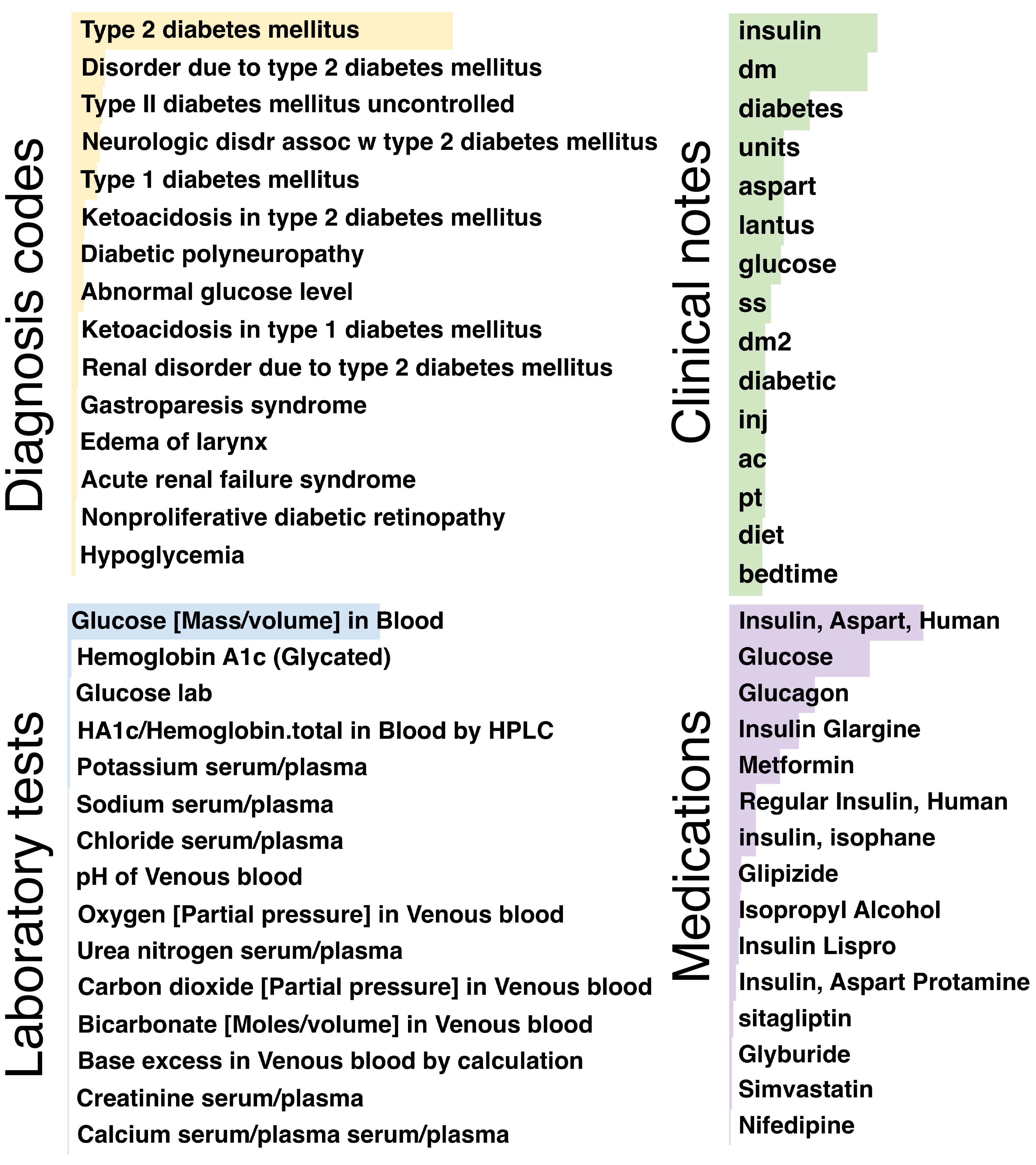}
	\caption{Example of learned phenotype and its probabilistic definition across the four data types (yellow for diagnosis codes, green for notes, purple for medications, and blue for laboratory tests). The mostly likely diagnosis code is assigned as label for the phenotype.)}
	\label{fig:ModelOutput1}
\end{figure}

\begin{figure}[h]
	\centering
	\includegraphics[width=\linewidth]{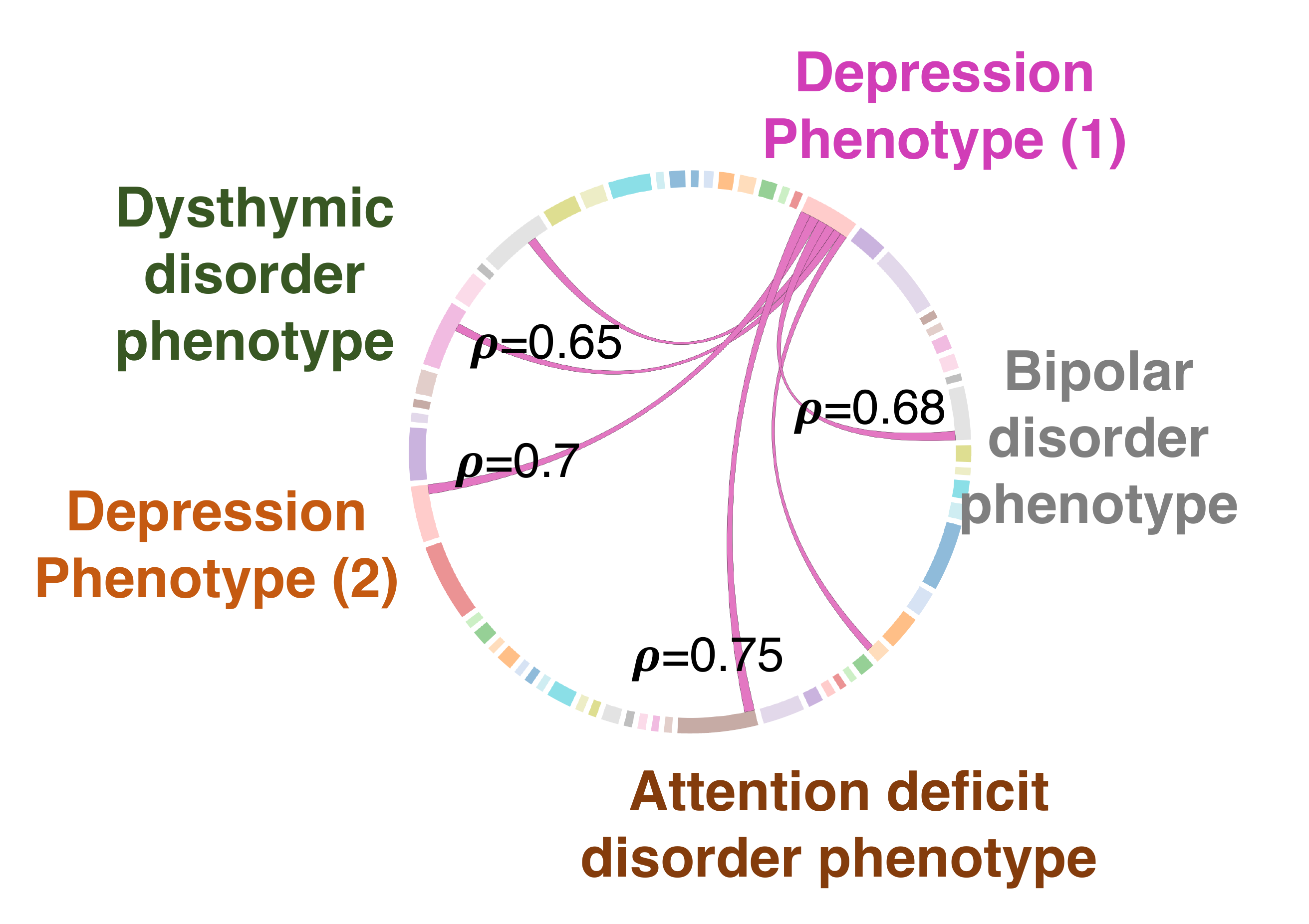}
	\caption{Example of five learned phenotypes and their learned correlations}
	\label{fig:ModelOutput2}
\end{figure}

\begin{figure}[h]
	\centering
	\includegraphics[width=\linewidth]{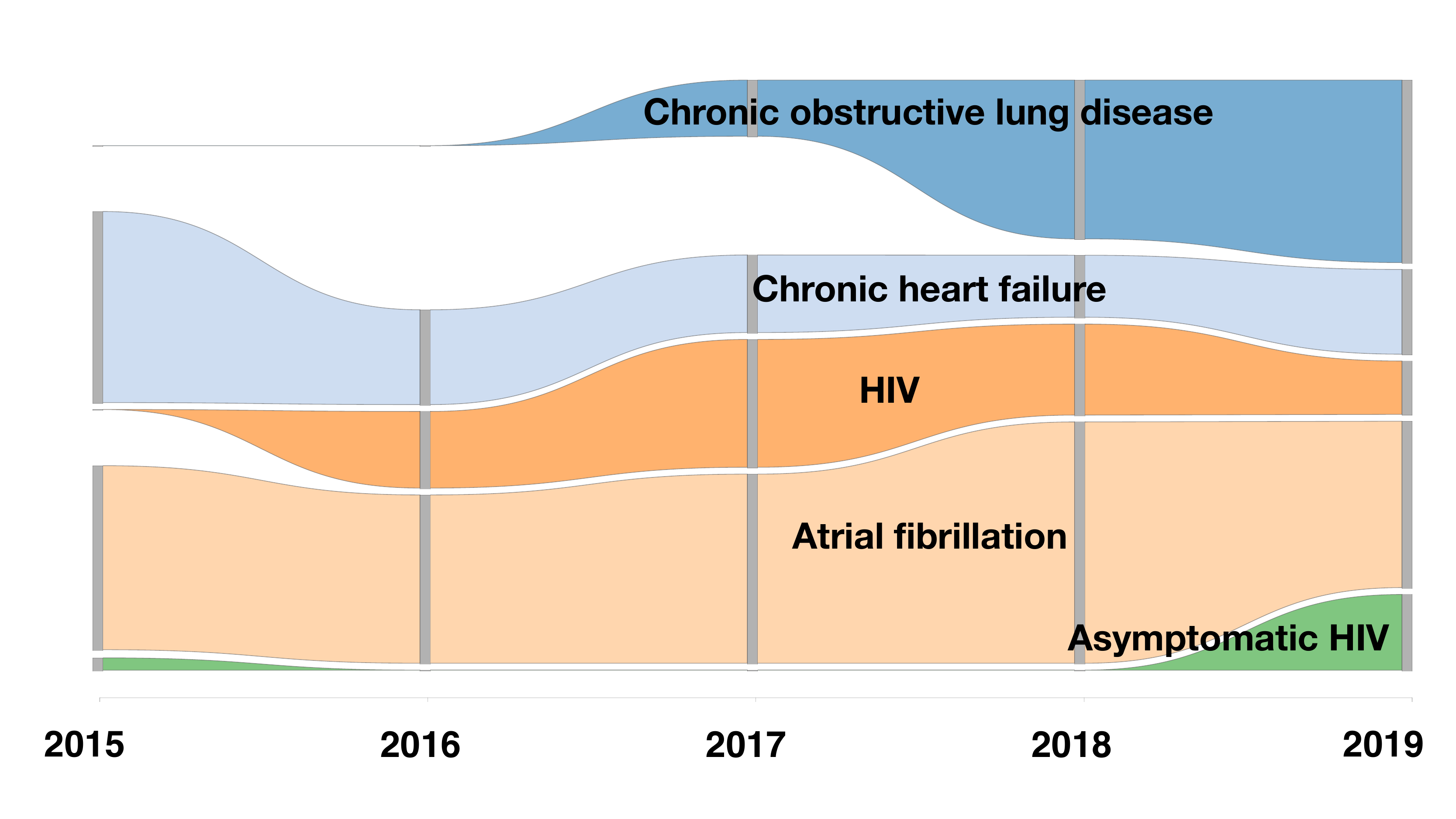}
	\caption{Example of patient-specific summary over five years. The top five most salient problems in 2019 are visualized and how their documentation has evolved through time. In this setup, the summary was produced at the year level by binning the patient's documentation for that time resolution. The patient has HIV-specific problems, although their HIV is becoming asymptomatic, as well as comorbidities, all cardiac in nature. (Relations among the inferred phenotypes are not shown. Dates are changed to maintain patient privacy.)}
	\label{fig:summary}
\end{figure}

To identify the appropriate granularity of the learned phenotypes, the model is trained on a target patient population of the same clinic. Each phenotype definition is composed of diagnoses, medications, laboratory tests, and clinical notes that have been observed to commonly co-occur in the training patient population. Figure~\ref{fig:ModelOutput1} shows an example phenotype learned by the model.   Figure~\ref{fig:ModelOutput2} shows phenotype-phenotype correlations learned by the model. Phenotypes are labeled with their most probable diagnosis code. The learned phenotypes from the model are then used to summarize a single patient EHR data over time. Figure~\ref{fig:summary} shows an automatically generated example summary of a single patient record over a five year period that leverages the proposed approach. \\

In this work we focus on phenotyping a population of patients from an HIV clinic in a large urban care institution. HIV-positive patient experience a high burden of disease, with many comorbidities due to the inflammatory nature of the virus and the toxicity of their medications \cite{gallant_Comorbidities_2017}. Moreover, due to high healthcare utilization, patients have complex and long medical histories that are difficult to sift through, exacerbating the need for summarization. We hypothesize that \textbf{1)} the model will learn many clinically valid phenotypes and phenotype relationships; \textbf{2)} training the model on an HIV-positive population will result in the identification of several HIV phenotypes, representing the different presentations and progression stages of HIV---a granularity that would likely be missed if trained on a more general and heterogeneous patient population; \textbf{3)} the model will also learn non-HIV phenotypes, representative of the many comorbidities of HIV; \textbf{4)} the model will identify correlations among phenotypes that indicate clinically valid relations of different types beyond simple is-a relationships.

\section{Methods}

\subsection{The model}
The model we proposed is based on the correlated topic model (CTM) \cite{blei_Correlated_2007}. In our context, topics are equivalent to phenotypes and documents are the patient records. We make a methodological contribution by expanding the CTM to support multiple input sources, beyond the single input source usually assumed in topic modeling. We make this important expansion to the model since unlike topic identification in general text, clinical documentation is more than just clinical notes. Instead, our model is able to learn phenotype definitions through identifying co-occuring patterns in clinical notes, laboratory tests, ordered medications, and diagnosis codes. Incorporating multiple sources of data into the phenotypes definitions allows for more robust phenotype definitions that can help overcome the inaccuracies present in just relaying on an single source of patient data. This is supported by previous work that has shown that incorporating heterogenous data yields superior phenotypes \cite{pivovarov_Learning_2015}. 

It has been previously proposed to leverage topic-model like models to learn clinical phenotypes. Our model differs in that we do not assume that phenotypes identified in each patient record are independent from one another. We remove the assumption of independence by allowing for phenotypes to be correlated. To do this our model, like the original CTM, replaces the traditional Dirichlet distribution used in Latent Dirchelet Allocation (LDA) \cite{blei_Latent_2003} to govern topic proportions with a logit-normal distribution\cite{blei_Correlated_2007}. The logit-normal distribution allows for phenotype proportions in each patient record to be correlated with one another (through the normal covariance matrix) but also sum up to 1 or 100\% of the patient record, as desired when modeling proportions. Changing the previously assumed Dirchelet distribution with a logit-normal distribution removes the conditional conjugacy between the posterior distribution and prior distribution of the phenotype proportions. To perform posterior inference Wang and Blei \cite{wang_Variational_2013} propose Laplace variational inference, a generalized form of variational inference that can handle non-conjugate models. In this paper we generalize the proposed Laplace Variational Inference even further to allow for multiple input types. This makes the model inference especially relevant to clinical data which contains many different data types. The model training is time-agnostic and treats each patient record as bag of observations, one for each data type. While motivation behind the model is to assign phenotypes on a single patient level for patient-level summarization, in this paper we focus on the learned phenotypes on the population level. Each phenotype is labeled using the most probably diagnosis code. 

\begin{figure}[!htbp]
	\centering
	\includegraphics[width=\linewidth]{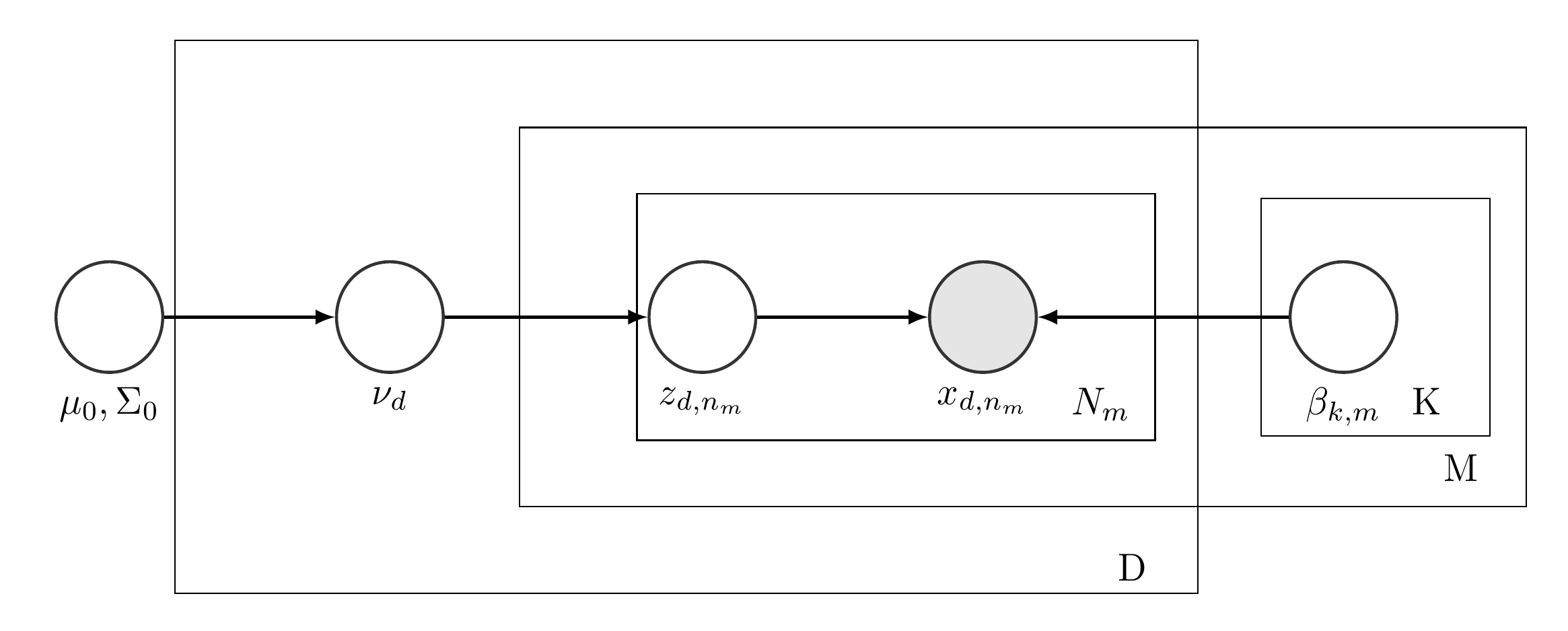}
	\caption{The graphical representation of the multi-input correlated topic model. Multiple inputs are represented by the additional plate notation M that is not present in the single-input CTM model. }
	\label{fig:model}
\end{figure}

The generative process of each patient record (D) with $N_m$ number of tokens for $M$ data types is provided below. The graphical representation of the model is presented in Figure~\ref{fig:model}.

\begin{enumerate}
	\item Draw log phenotype proportion $\nu_d \sim N(\mu_0, \Sigma_0)$
	\item For each $n_m$ token ($x_{d,n_m}$) in data type ($m=1,...,M$):
	\begin{enumerate}
		\item Draw phenotype assignment $z_{d,n_m}|\nu_d \sim Mult(\pi(\nu_d))$
		\item Draw token $x_{n_m}|z_{n_m}, \beta_{k,m} \sim Mult(\beta_{z_{n_{m}}})$
	\end{enumerate}	
\end{enumerate}

\subsection{Probabilistic inference}

The phenotype definitions and their correlations with one another are obtained through performing Bayesian posterior inference which estimates the conditional probability of the unobserved or latent model variables given the observed model variables. In the case of the proposed model this means calculating the probability of the phenotype proportions of each patient record ($\nu$) and phenotype assignment of each input ($Z_{nm}$) given the observed patient data ($X_{nmd}$) and phenotype distributions ($\beta$), or $p(\nu,z_{}| x_{d,n_m}, \beta)$. When the posterior distribution has a conjugate prior this greatly simplifies the Bayesian analysis and allows for the use of popular sampling methods for approximate inference such as Markov chain Monte Carlo sampling such as Gibbs sampling as employed in \cite{pivovarov_Learning_2015}. 

However, conjugacy limits the types of distributions used in the model, and thus restricts the flexibility of data modeling. In order to allow for phenotypes to be correlated with one another the prior distribution used to model the phenotype proportions in the patient record needs to allow for phenotype correlations. Since that is not possible with the Dirchelet distribution, it needs to be replaced with a different distribution that meets this criteria. However since the Dirichlet distribution is the conjugate prior to the multivariate distribution used to model the phenotype data assignments, this modeling change means that the model losses its conditional  conjugacy. Hence, deterministic approximate inference methods such as variational inference is more feasible than other sampling methods. 

By contrast to sampling approximation methods for inference, the theoretical guaranties of convergence of variational inference methods to the true posterior have been less studied. However, variational inference has become a popular inference method in Bayesian statistics as it tends to be faster and scale better with large and complex data \cite{blei_Variational_2017}. Even in variational inference, some popular implementations such as mean-field variational depend of conjugate models. Wang and Blei \cite{wang_Variational_2013} propose Laplace variational inference, a generalized form of variational inference that can handle non-conjugate models. The method uses Laplace approximations in the coordinate ascent updates within the variational optimization problem. This methods was shown to generalize to different types of non-conjugate model and have superior performance compared to the original ad-hoc inference method previously proposed here \cite{blei_Correlated_2007}.  In this paper we generalized the Laplace Variational Inference for multiple input types. The mathematical derivation of the Laplace variational inference with multiple input types is shown below.

\subsection{Inference mathematical derivation}

As presented in the graphical model (see Figure~\ref{fig:model}), the under-script $m$ represents the $m$-th input type, where $m=1,...,M$.  The derivation below contributed to the previously proposed inference by  \cite{wang_Variational_2013} by allowing for M input types instead of a single input type. The model is represented by the joint probability distribution in equation (1). The inference problem is to solve for the posterior distribution which is the conditional distribution of the latent variables $\nu$ and $z$ given $x$ in equation (2). 

\begin{equation}
p(\nu, z, x)= \prod_{m=1}^{M}p(x_m|z_m)p(z_m|\nu)p(\nu) 
\end{equation}

\begin{equation}
p(\nu, z|x)=\frac{p(\nu, z, x)}{\int p(\nu,z,x)dz d\nu}
\end{equation}

The integral in the denominator of equation (2) is intractable to compute exactly \cite{blei_Correlated_2007}. 
As proposed by \cite{wang_Variational_2013} the posterior is approximated using Laplace Variational Inference through optimization. A family of densities are posited over the latent variables. The model assumptions include: 

\begin{enumerate}
	\item The variational distribution is fully factorized:
	
	\begin{equation}
	q(\nu, z)=q(\nu)\prod_{m=1}^{M}q(z_m)
	\end{equation}

	\item $\nu$ is real valued and $p(\nu)$ is twice differentiable with respect to $\nu$
	
	\item The distribution $p(z_m|\nu)$ is in the exponential family:
	
	\begin{equation}
	p(z_m|\nu)=h(z_m)exp\{\eta(\nu)^Tt(z_m)-a(\eta(\nu))\}
	\end{equation}
	
	\item The distribution $p(x_m|z_m)$ is in the exponential exponential family such that:
	
	\begin{equation}
	p(x_m|z_m)=h(x_m)exp\{t(z_m)^T<t(x_m),1>\}
	\end{equation}
	
\end{enumerate}

In variational inference the approximation for the posterior distribution is obtained through minimizing the Kullback-Leiber (KL) divergence to the exact posterior. 

\begin{equation}
q^*(\nu,z)=argmin KL(Q(\nu, z) || P(\nu, z|x)
\end{equation}

Under standard variational inference theory minimizing the KL divergence between $q(\nu,z)$ and the true posterior $p(\nu,z|x)$ is the same as maximizing the lower bound of the log marginal likelihood of observed data $x$. Using Jensens's inequality the variational object $L(q)$ is defined by equation (5). 

\begin{equation}
\begin{aligned}
log p(x) & = log \int p(\nu,z,x)dz d\nu \\
& \ge E_q[log (p(\nu, z, x))] - E_q[log(q(\nu,z))]  \\
& = E_q[log (p(\nu, z, x))] - E_q[log(q(\nu)\prod_{m=1}^{M}q(z_m)]  \\
& \equiv L(q)
\end{aligned}
\end{equation}

Setting the partial derivative of $L(q)$ with respect to $q$ to zero provides the optimal variational updates to $q(\nu)$ and $q(z_m)$ seen in Equations (9) and (10). When $p(\nu)$ is conjugate to $p(z_m|\nu)$ then equations (5) and (6) have closed form solutions. In the case of this non-conjugate model \cite{wang_Variational_2013} put forward approximates to the updates using Laplace approximation. 

\begin{equation}
q^*(\nu) \propto exp \{ E_{q(z)}[log \prod_{m=1}^{M} p(z_m|\nu)p(\nu)]\}
\end{equation}

\begin{equation}
\begin{aligned}
q^*(z_1) &\propto exp \{ E_{q(\nu)}[log p(x_1|z_1)p(z_1|\nu)]\} \\
&\vdots \\
q^*(z_m) &\propto exp \{ E_{q(\nu)}[log p(x_m|z_1)p(z_m|\nu)]\}
\end{aligned}
\end{equation}

The following is the derivation of the variational update to $q^*(\nu)$ using the previously stated assumption that $p(z_m|\nu)$ is assumed to belong to the exponential family.

\begin{equation}
\begin{aligned}
q^*(\nu) &\propto exp \{E_{q(z)}[log\prod_{m=1}^Mp(z_m|\nu)p(\nu)]\} \\
&=exp\{E_{q(z)}[logp(\nu)+\sum_{m=1}^Mlog(z_m|\nu)]\} \\
&=exp\{E_{q(z)}[\sum_{m=1}^M log (h(z_m)exp\{\eta(\nu)^Tt(z_m)-a(\eta(\nu))\})) +log p(\nu)]\} \\
&= exp\{E_{q(z)}[\sum_{m=1}^M (\eta(\nu)^Tt(z_m)-a(\eta(\nu)) ) +logp(\nu) ] \} \\
&=exp\{E_{q(z)}f(\nu)\}
\end{aligned}
\end{equation}

The function $f(\nu)$ in Equation (10) has no closed form and this is approximated with the following 2nd order Taylor approximation around $\hat{\nu}$ which is the $\nu$ that maximizes $\nabla f(\nu)$.

\begin{equation}
f(\nu)\approx f(\hat{\nu})+ \nabla f(\hat{\nu})(\nu - \hat{\nu}) + \frac{1}{2}(\nu - \hat{\nu})^T\nabla^2 f(\hat{\nu})(\nu - \hat{\nu})
\end{equation}

Thus the update for $q(\nu)$ is approximate with $\mathcal{N}(\hat{\nu},-\nabla^2f(\hat{\nu})^{-1}) $

The sufficient statistics of the exponential family are: 

\begin{equation}
\begin{aligned}
h(z_m)&=1 \\
t(z_m)&=\sum_n z_{mn} \\
\eta(\nu)&=\nu - log \{\sum_k exp \{ \nu \} \\
a(\eta(\nu))&=0
\end{aligned}
\end{equation}

Using the sufficient statistics above $f(\nu)$ is the following:

\begin{equation}
\begin{aligned}
f(\nu) &= \sum_{m=1}^M (\eta(\nu)^T - a(\eta(\nu))-\frac{1}{2} (\nu - \mu_0)^T\Sigma_0^{-1}(\nu-\mu_0) \\
&= \eta(\nu)^T\sum_{m=1}^M \{E_{q(z)}[t(z_m)]\} - \frac{1}{2}(\nu-\mu_0)^T\Sigma_0^{-1}(\nu-\mu_0)
\end{aligned}
\end{equation}

The first derivative and second derivative of $f(\nu)$ are the following: 

\begin{equation}
\begin{aligned}
\nabla f(\nu) &= \pi_i (1_{[i=j]}- \pi_j)\sum_{m=1}^M E_{q(z)}[t(z_m)]-\Sigma_0^{-1}(\nu - \mu_0) \\
&=\sum_{m=1}^M E_{q(z)}[t(z_m]- \pi\sum_{k=1}^K[\sum_{m=1}^ME_{q(z)}[t(z_m)]]_k - \Sigma_0^{-1}(\nu -\mu_0)
\end{aligned}
\end{equation}

where $\pi \propto exp\{\eta(\nu)\}$ 

\begin{equation}
\nabla^2 f(\nu)_{ij} = (-\pi_i 1_{i=j}+ \pi_i \pi_j)\sum_{k=1}^K[\sum_{m=1}^M E_{q(z)}[t(z)]]_k - (\Sigma_0^{-1})_{ij}
\end{equation}

The update to $q(z_m)$ where $m= 1,...,m$ is the following:  

\begin{equation}
\begin{aligned}
q^*(z_m) &\propto exp \{ E_{q(\nu)} [log p(x_m |z_m)p(z_m|\nu)] \} \\
&= exp \{ log p(x_m |z_m)+ E_{q(\nu)}[log p(z_m|\nu)]\}
\end{aligned}
\end{equation}

Using the exponential form of $p(z_m|\nu)$ and $p(x_m|z_m)$:

\begin{equation}
\begin{aligned}
log q(z_m)&= log p(x_m| z_m)+ E_{q(\nu)}[log p(z_m|\nu)] \\
&=log p(x_m|z_m)+log h(z_m)+ E_{q(\nu)}[\eta(\nu)^T]t(z_m)- C \\
&=log h(z_m) + t(z_m)^T<t(x_m), 1> + log h(z_m) + \\
&  E_{q(\nu)}[\eta(\nu)^T]t(z_m)-C
\end{aligned}
\end{equation}

\begin{equation}
q(z_m)\propto h(z_m)exp\{E_{q(\nu)}[\eta(\nu)^T]+ t(x_m)^Tt(z_m)\}
\end{equation}


\subsection{Dataset} 
The model was trained on the EHR data of 7,523 patients from an HIV clinical from a large urban care institute. Patient data used was fully identified for which the use was approved by the Institutional Review Board of our institution. The data spanned 8 years and included the data types: words from clinical notes, laboratory tests ordered, medication orders, and assigned diagnoses codes from across all clinical settings (inpatient, outpatient, emergency). For the purpose of the model training each patient record was restricted to the most recent 2.5 years data. The final training dataset included the following total data counts and unique vocabulary size in brackets: total words from clinical notes: 128,034,516 (unique: 25,894); total laboratory tests: 463,524 (unique: 129); total medications: 510,820 (unique: 6,714); and total diagnosis codes: 246,623 (unique: 2,956). 

\subsection{Model training and parameter selection}

The parameters of the normal distribution governing phenotype proportions $\nu_d$ were initialized with $\mu_0$ equal to a zero vector and $\Sigma_0$  set to the identity matrix. The phenotype distribution $\beta_{k,m}$ for each input type was initialized with a Uniform distribution over the (K-1) simplex. This equivalent to initializing topics with a Dirichlet distribution with parametrization of 1.  A small amount of random positive noise was added to each uniform distribution so there was a small variation in the initial phenotypes. Three alternatives of the model were estimated (K=50, 100, 250). 

To identify the best performing model of the three alternative number of phenotype (K=50, 100, 250), a clinical expert reviewed 20 randomly selected phenotypes from each model. The best performing model is further evaluated for the clinical correctness of the phenotypes and phenotype-relatedness learned by the model.

\section{Evaluation Setup}

We evaluate our hypotheses 1 through 4 using a mixture of qualitative and quantitative evaluations. Qualitative evaluation of the phenotypes and phenotype relatedness was performed by two clinical experts. The quantitative evaluation was performed through a comparison to the Clinical Classification Software (CCS), which provides expert-curated manual classification of diagnosis codes into largely clinically homogeneous groups \cite{agencyforhealthcareresearchandquality_HCUP_2017}

\subsection{Hypothesis 1: clinical validity}
To evaluate the clinical validity of the learned phenotypes, 50 randomly selected phenotypes were evaluated independently by two clinicians. The phenotypes were evaluated according to their coherence, granularity, and label quality \cite{pivovarov_Learning_2015}. Previous works citing clinical evaluation of phenotypes by experts have reported the scoring of a single clinician \cite{pivovarov_Learning_2015, ho_Limestone_2014}. Since this scoring can very subjective, we opted for two clinicians to score the phenotypes and the final score assigned is the average of the two clinicians. Since we did not want the opinion of one clinicians to be influenced by the other, there was not adjudication stage in the scoring (common on qualitative rating tasks made by more than one reviewer). This made the qualitative evaluation a very stringent task. We provide an analysis of the agreement between the clinicians scoring which can illuminate the level of subjectivity of this type of evaluation.

\paragraph{Phenotype coherence.}
Phenotype coherence is meant to capture the quality of each learned phenotype according to its most probably observations. A coherent phenotype is defined to describe a single condition with few or no unrelated observations (clinical words, labratory tests, medications, and diagnosis codes). The expert was asked to rate each phenotype as having: `bad coherence' (score=1) , `some coherence' (score=2), `good coherence' (score=3), or `excellent coherence' (score=4).  Phenotypes with `bad coherence' should look like a random combination of observations, `some coherence' indicates the observations assigned to the phenotype are somewhat related to one another, `good coherence' indicates the phenotype is a very good representation of a disease, and `excellent coherence' indicates the phenotype definition has almost no unrelated observations assigned to it. 

\paragraph{Phenotype granularity.}
The clinical experts were asked to characterize the granularity of each randomly selected phenotype by assessing whether the model learned a `single disease' (score=3), a `group of diseases' (score=2), or a 'non-disease' phenotype (score=1).  

\paragraph{Label quality.}
The representativeness of the automatically assigned phenotype label of the phenotype as a whole was evaluated. Each label was categorized by the clinical experts as `unrelated' to the rest of the phenotype (score=1),  `related' to the rest of the phenotype (score=2), or `actionable' (score=3).  Labels that were deemed as actionable are those representative of a single phenotype and have the appropriate granularity to provide a clinician information that could be used without additional information to guide further testing, diagnosis, or counseling.  \\

\paragraph{Phenotype relatedness.}

Next, the clinical validity of the phenotypes-relatedness were evaluated by a single clinical expert. The expert reviewed all phenotypes relationships that were indicated to have a correlation greater than 0.5 correlation coefficient. Two sets of phenotype-relationships evaluated:  1) positive phenotype relationships learned between "more common" non-hiv phenotypes, defined as phenotypes that were represented in more than 5\% of the patient population in our dataset; and 2) positive relationships learned between "rarer" non-hiv phenotypes, represented in 5\% of sample population or less. The justification for evaluating relationships between "more common" phenotypes is that the model findings are grounded in more patient record, which could result in more robust findings. However, evaluating phenotype relationships identified between "rarer" phenotypes could still be interesting to assess in case the model is able to identify less known clinical relationships.

\subsection{Hypothesis 2: focus on HIV phenotypes}

Our second hypothesis was that since HIV is a complex disorder with diverse presentations and severity among patients, the model would identify several distinct HIV phenotypes. 
In evaluating this hypothesis, we wished to understand to what extent the model is able to learn multiple clinically valid HIV phenotypes and also characterize what those phenotypes were. To do so  we had an HIV clinical expert review all the phenotypes automatically labeled as 'HIV'. The clinical expert was asked to i) indicate if the phenotype was clinically valid, ii) indicate if the phenotype was indeed an 'HIV' phenotypes; and iii) give a more granular description of the phenotype if it was indeed an 'HIV' phenotype in order to assess if the model identified disease progression, presentation, or acuity.

\subsection{Hypothesis 3: focus on non-HIV phenotypes}

To assess if the model was able to learn diverse phenotypes, representative of the many comorbidities of HIV we quantitatively compared the phenotypes learned to the disease groups identified in the CCS. We did this by categorizing all 250 learned phenotypes according to their labels' corresponding CCS level-1 category. If the model was able to learn phenotypes that fit into many CCS categories, we would could conclude that the model was able to learn diverse types of phenotypes, beyond HIV.

\subsection{Hypothesis 4: types of phenotype- relatedness }

We performed two evaluations to assess whether the model was able to identify correlations among phenotypes that indicate clinically valid relations of different types beyond simple is-a relationships. The first evaluation included a clinical expert review phenotypes identified by the model as highly related and determine what kind of relationship type the model learned. Example relationship types include comorbidities, same phenotype, phenotype subtype, and others.  The second evaluation we counted how many significant relations learned by the model indicated an is-a relationship, as evidenced by same level-1 CCS categories, versus a more diverse relation type such as comorbidity when spanning different CCS categories.

\section{Results}
The qualitative evaluation by the clinical expert indicated that the 250-phenotype model yielded the most coherent and granular phenotypes of the three models (K=50, 100, 250). All results below are described for the evaluation performed for the K=250 phenotype model.

\subsection{Hypothesis 1: clinical validity} 

\paragraph{Phenotype quality}

\begin{figure}[h]
	\centering
	\includegraphics[width=\linewidth]{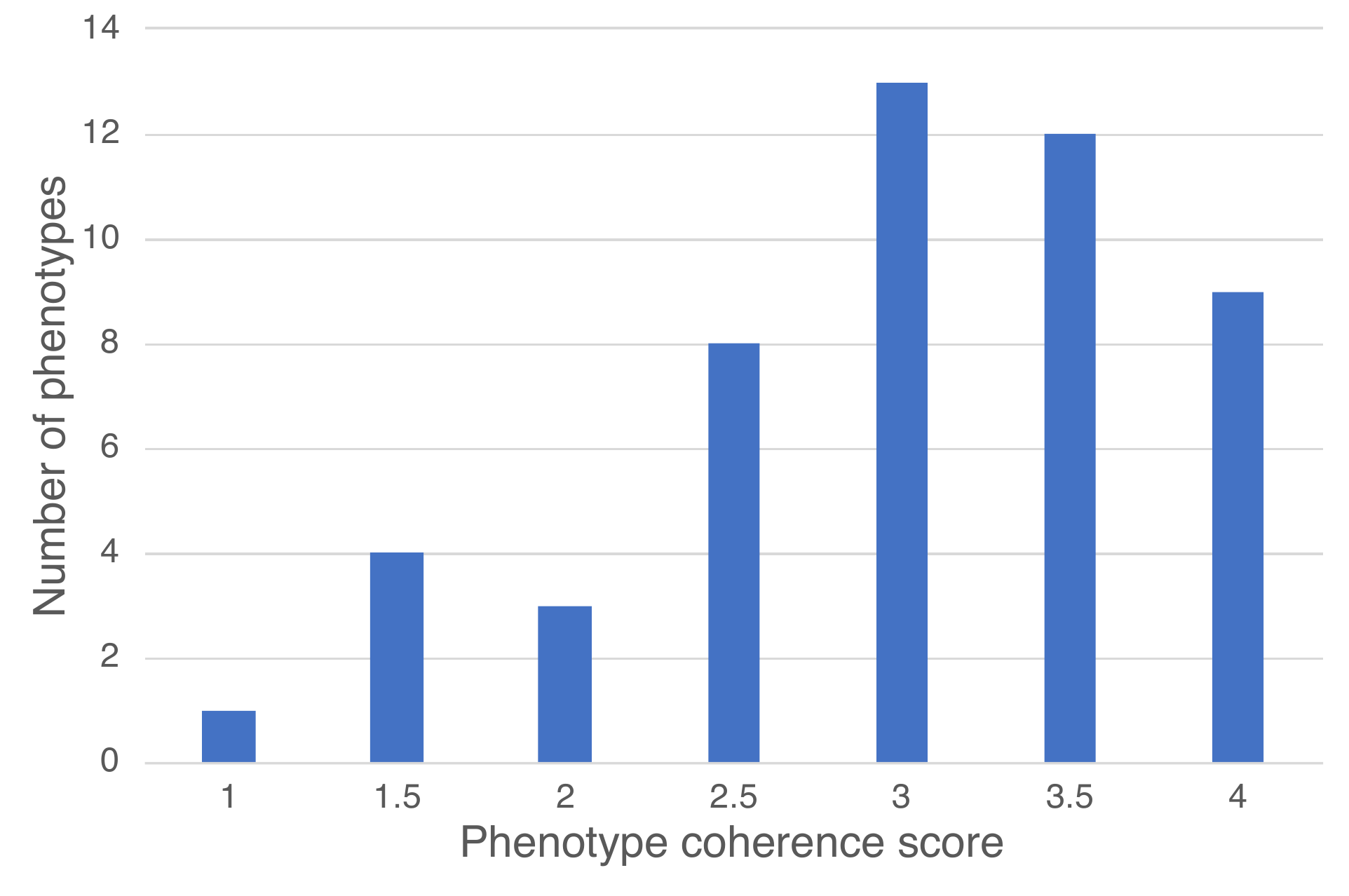}
	\caption{Phenotype coherence scores. Average score across the two clinical expert scores. Score 1=`bad coherence', 2=`good coherence', 3= `very good coherence', 4=`excellent coherence'.}
	\label{fig:coherence}
\end{figure}

\begin{figure}[h]
	\centering
	\includegraphics[width=\linewidth]{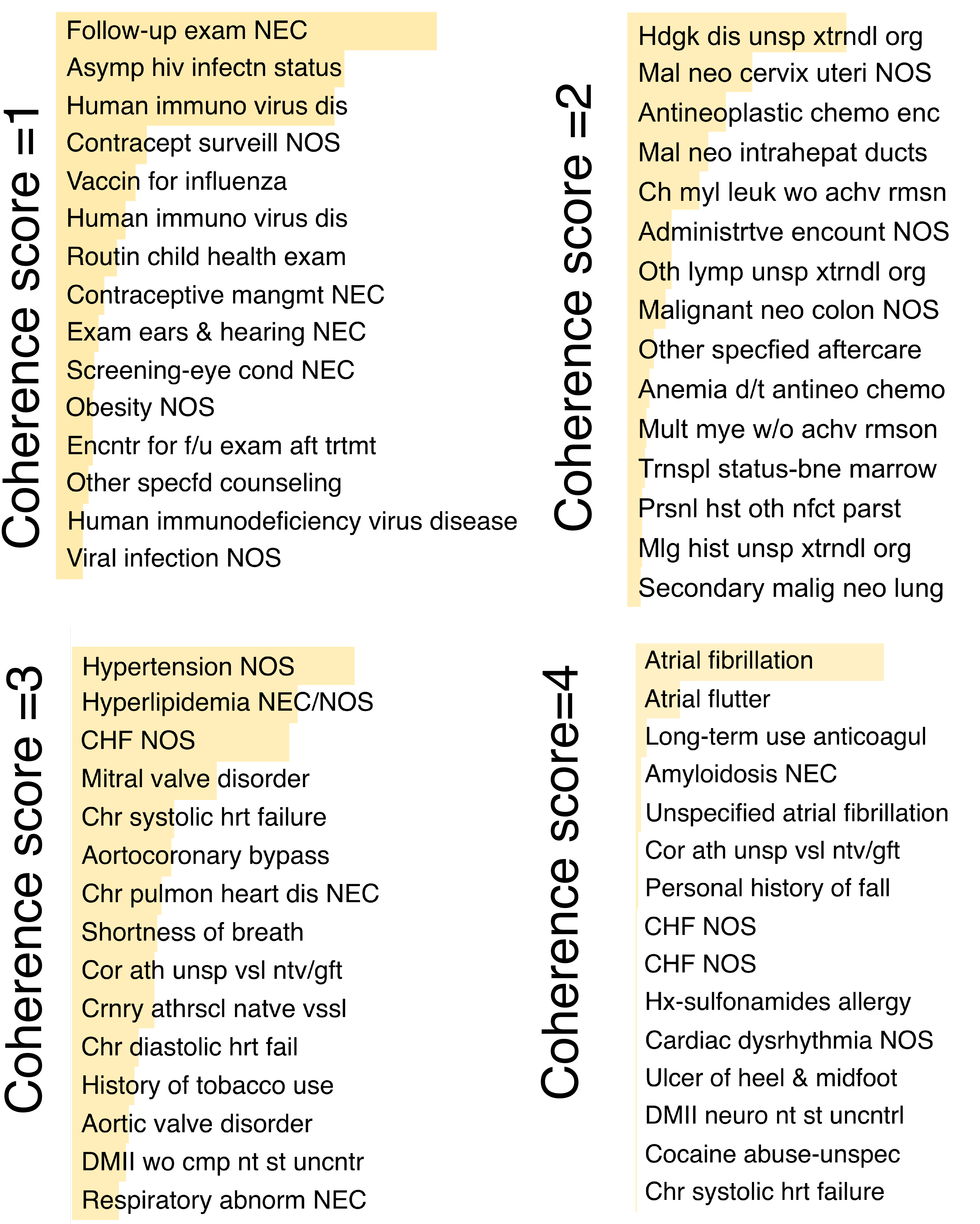}
	\caption{Example phenotypes by coherence score assigned by the clinical experts. Each phenotype is represented here by its top diagnosis codes rather than all 4 data types for the sake of space. Score 1=`bad coherence', 2=`good coherence', 3= `very good coherence', 4=`excellent coherence'.}
	\label{fig:coherence_examples}
\end{figure}

Of the 50 evaluated phenotypes from the 250-phenotype model, 10\% of the phenotypes (n=5) were deemed to have no coherence (average coherence score of 1 or 1.5) while the large majority of evaluated phenotypes (n=45) were deemed to be coherent (with average coherence score of 2 or above) (see Figure~\ref{fig:coherence}). The most number of phenotypes were scored as having 'good coherence' (n=13), followed by 12 phenotypes with an average of 3.5 (between `good coherence' and `excellent coherence'). The `bad coherence' phenotypes were found to be non-disease specific, but instead captured documentation related to general primary care visits. Figure~\ref{fig:coherence_examples} shows the diagnosis codes of example phenotypes with coherence scores 1 (`bad coherence') through 4  (`excellent coherence') by both the clinical experts. The phenotypes in the example identified a clinic visit phenotype (scored 1), grouping of cancers phentoype (scored 2),  grouping of heart diseases phenotype (scored 3), and an Aterial fibrillation phenotype (scored 4). 


Comparing the coherence phenotype scoring assigned by the two clinicians we found that the two clinicians had a low agreement on the exact coherence score assigned to the phenotypes (scores 1 through 4) but that the average difference between the scores was less than 1 point (0.9). This indicates that the clinicians evaluation of the phenotypes was not far apart. When comparing the clinician agreement on whether a phenotype was identified as not coherent (score of 1) versus coherent (score of 2 and above) the agreement was high, at 90\% of the evaluated phenotypes (see Table ~\ref{tab:agr}). Of the 5 phenotypes that the reviewers did not agree on, 4 looked like HIV clinic well visits. The disagreement seemed to stem from whether the model identified a disease phenotype or a clinical-settings phenotype.  An example such phenotype had the following top 5 diagnosis codes: `Human immune virus disease', `Obesity NOS', `Elevated blood pressure w/o hypertension', `Hypertension NOS', and `Laboratory exam NOS'.   

\begin{table}
	\caption{Comparison of 2 clinician scoring for phenotype coherence}
	\label{tab:agr}
	\begin{tabular}{lccc}
		\toprule
		&&\multicolumn{2}{c}{Clinician 1} \\
		&&Not Coherent &  Coherent \\
		\midrule
		Clinician 2 & Not Coherent & 1&1\\
		&Coherent &4&45\\
		\bottomrule
	\end{tabular}
\end{table}

\begin{figure}[h]
	\centering
	\includegraphics[width=\linewidth]{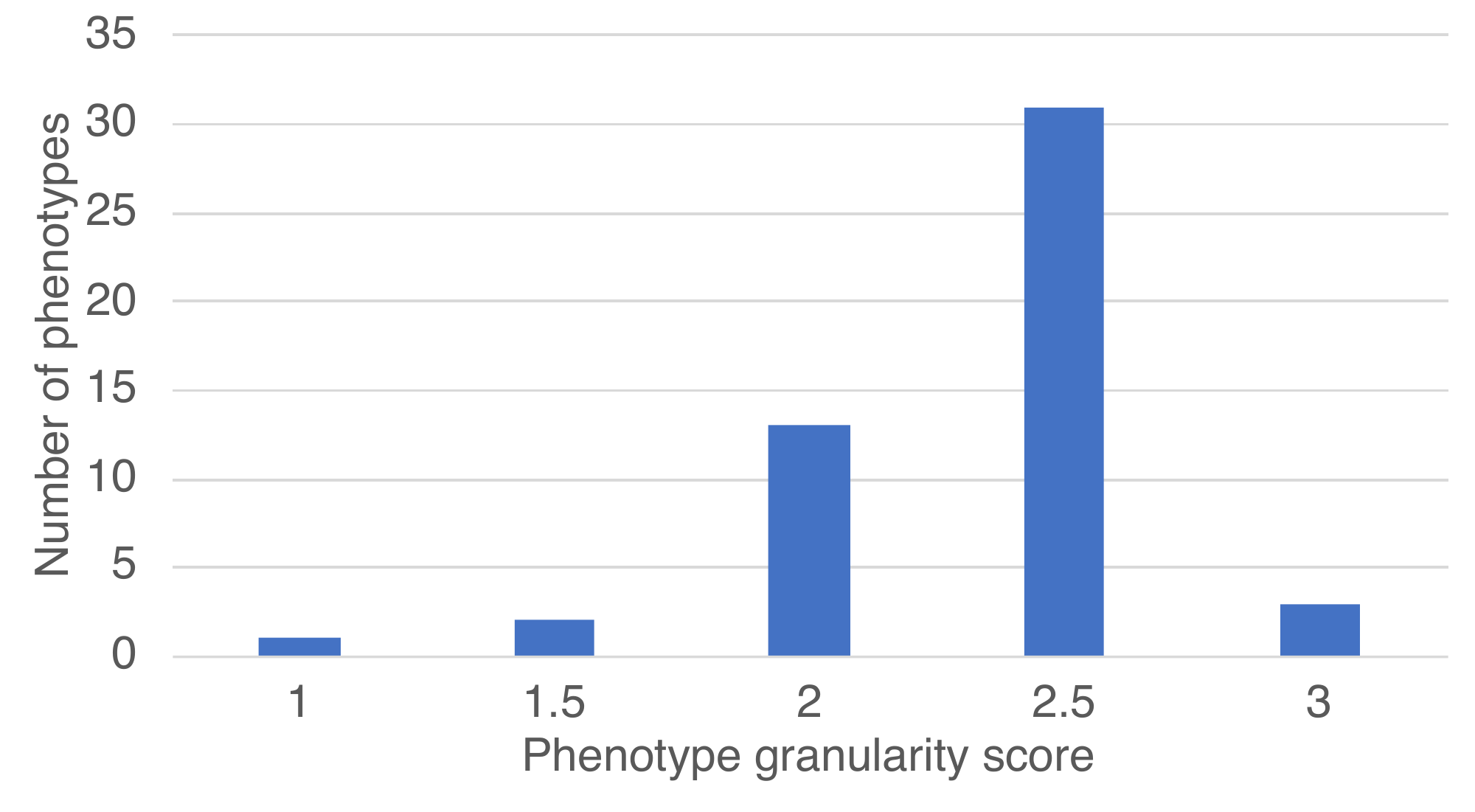}
	\caption{Phenotype granularity scores. Average score across the two clinical expert scores. Score 1='non disease', 2='group of diseases', 3= 'single disease'.}
	\label{fig:gran}
\end{figure}

The phenotype granularity scores indicated that 90\% of the evaluated phenotypes (n=47) had a granularity score 2 or greater (see Figure~\ref{fig:gran}). This means that almost all of the evaluated phenotypes were deemed by both reviewers to identify a single or a group of diseases. The most number of phenotypes were assigned an average score of 2.5  (n=31), the next most prevalent score was 2 (n=13). This indicates that the model mostly identified phenotypes that were a group of diseases rather than a single disease. An example of a phenotype that had an average granularity score of 2.5 had different diagnoses codes identifying a fall or accident and different body parts such as shoulder, forearm, limb and hand.  One reviewer scored the phenotype as identifying a single disease being `limb injury due to accident', while the other reviewer believed that since multiple body parts were identified the phenotype represented a group of diseases.  

\begin{figure}[h]
	\centering
	\includegraphics[width=\linewidth]{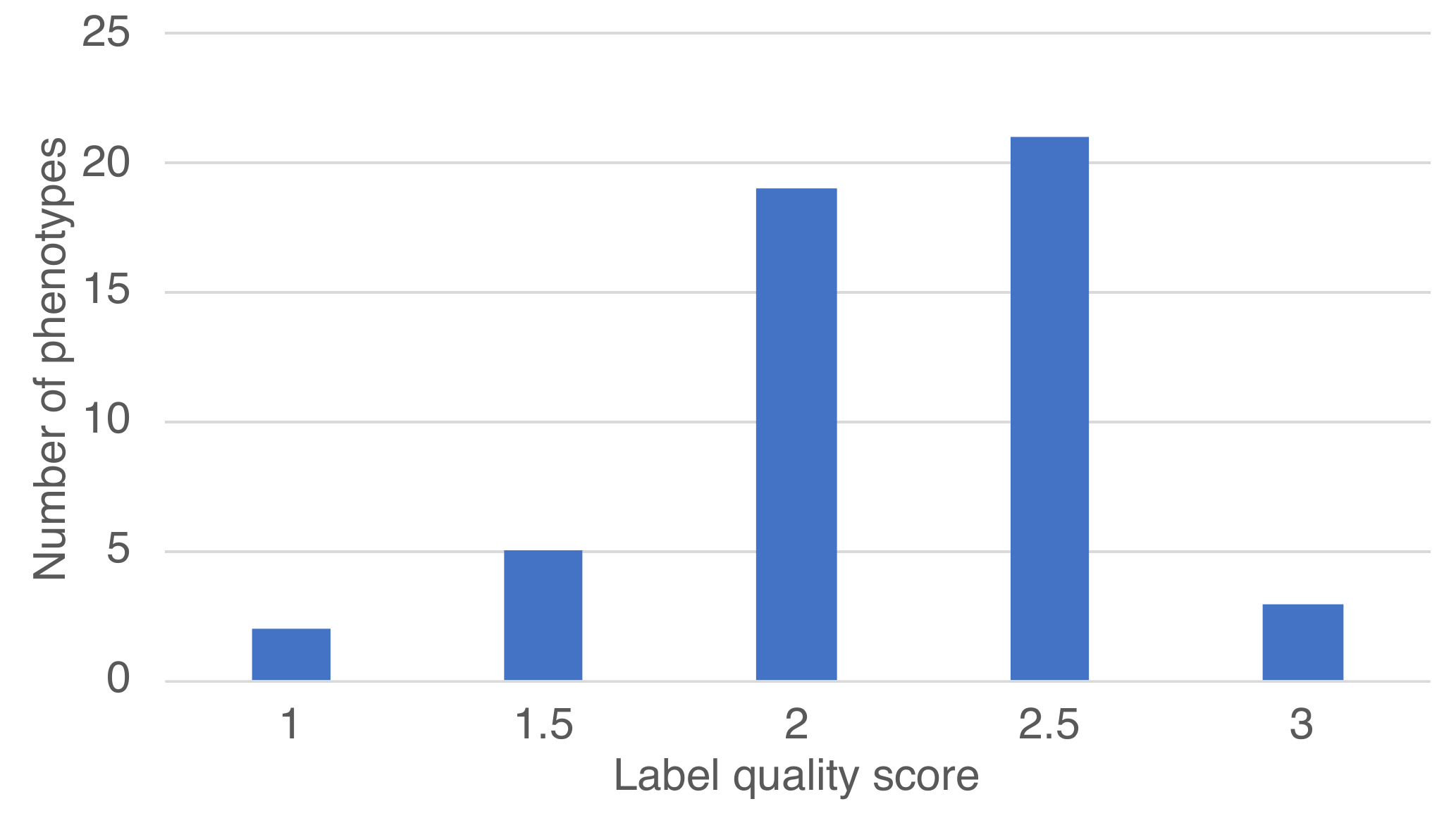}
	\caption{Phenotype coherence scores. Histogram of average phenotype coherence scores assigned by the two clinical expert. Score 1=`not related', 2=`related', 3= `actionable'.}
	\label{fig:label}
\end{figure}

 The phenotype labels were mostly found to be `related' with a score of 2.5 (n=21) or 2 (n=19) (see Figure~\ref{fig:label}). Only 3 phenotype labels were identified as `actionable' with a score of 3 by both reviewers. The feedback from the reviewers was that the diagnosis code used for the phenotypes was too granular to adequately represent the entire phenotype.

%

\begin{figure}[h]
	\centering
	\includegraphics[width=\linewidth]{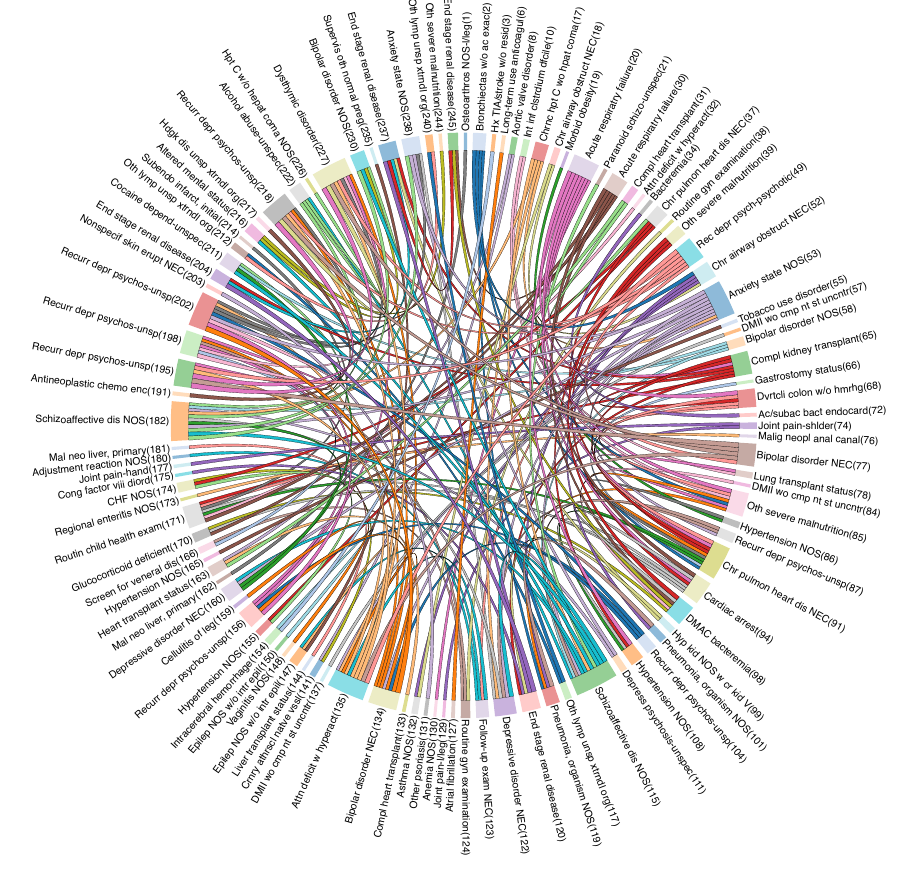}
	\caption{All significant pairwise-positive correlations visualized}
	\label{fig:all_positive_corr}
\end{figure}

\paragraph{Phenotype-relatedness quality}
Of the learned phenotype-pair correlations, 471 (1.5\% of all possible phenotype-phenotype pairs) were significant (correlation coefficient above 0.5 in absolute value).  Of the 471 significantly correlated phenotype-pairs, 395 where positive correlated (Figure~\ref{fig:all_positive_corr}) and 76 were negatively correlated. We had a clinical expert perform clinical validity of the learned phenotype relationships. 

\begin{figure}[h]
	\centering
	\includegraphics[width=\linewidth]{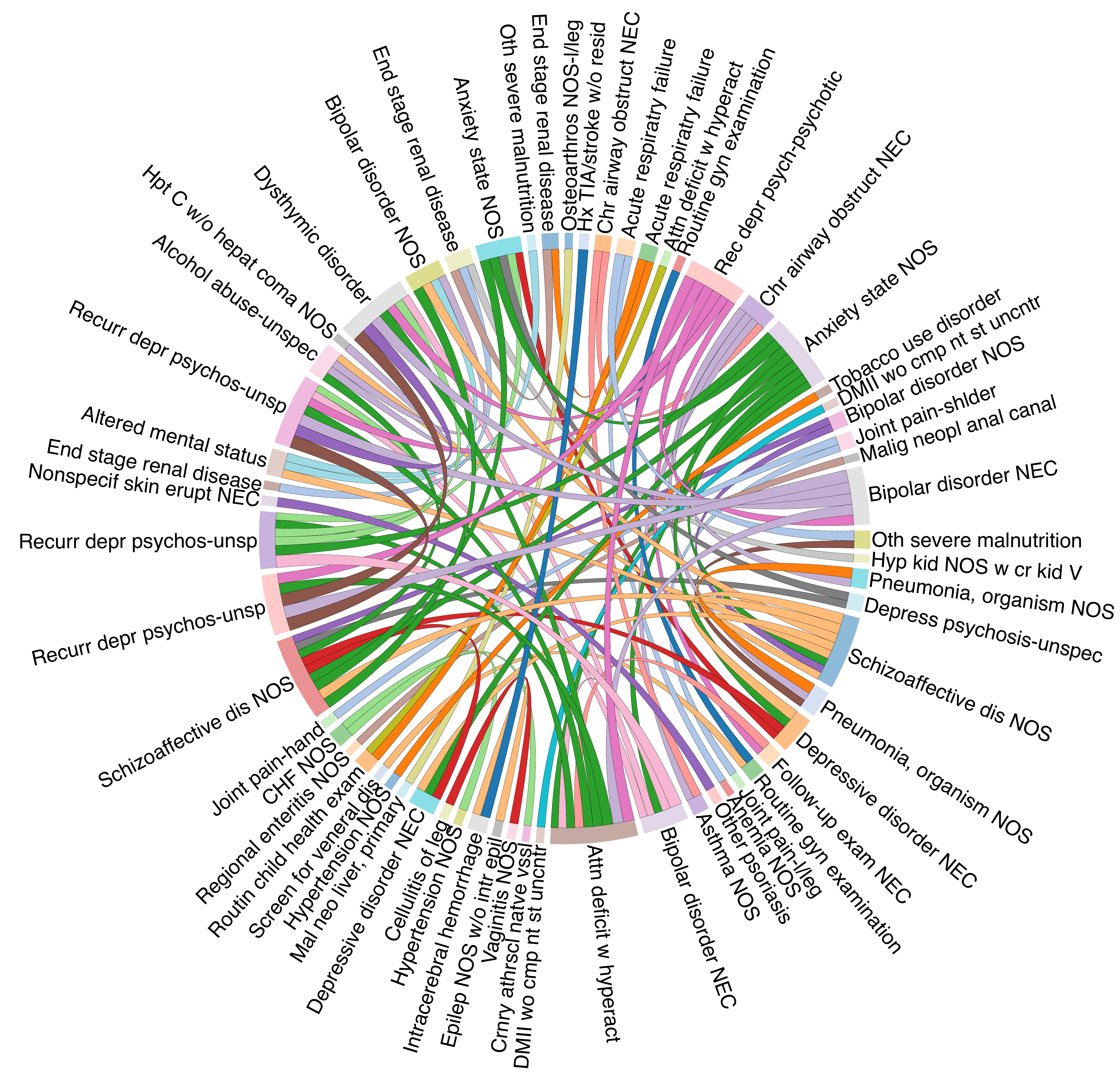}
	\caption{Significant pairwise-positive correlations evaluated by clinician for clinical correctness.}
	\label{fig:more_than5}
\end{figure}

In the "more common" phenotype set, 82 phenotype pairs were found to have a correlation greater than 0.5 (Figure~\ref{fig:more_than5}). These 82 correlations resulted from 61 unique phenotypes, hence on average each phenotype had more than significant correlation with more than one phenotype. Of the 82 reviewed relations 80 (98\%) were found clinically valid. One relation rated non clinically valid was the high correlation between a non-disease phenotype for outpatient visits and a non-disease phenotype for  inpatient visits. The other non clinically valid relation was between a joint disease phenotype and a phenotype that seemed to be a mix of hepatitis C, liver disease, and obesity.

\begin{figure}[h]
	\centering
	\includegraphics[width=\linewidth]{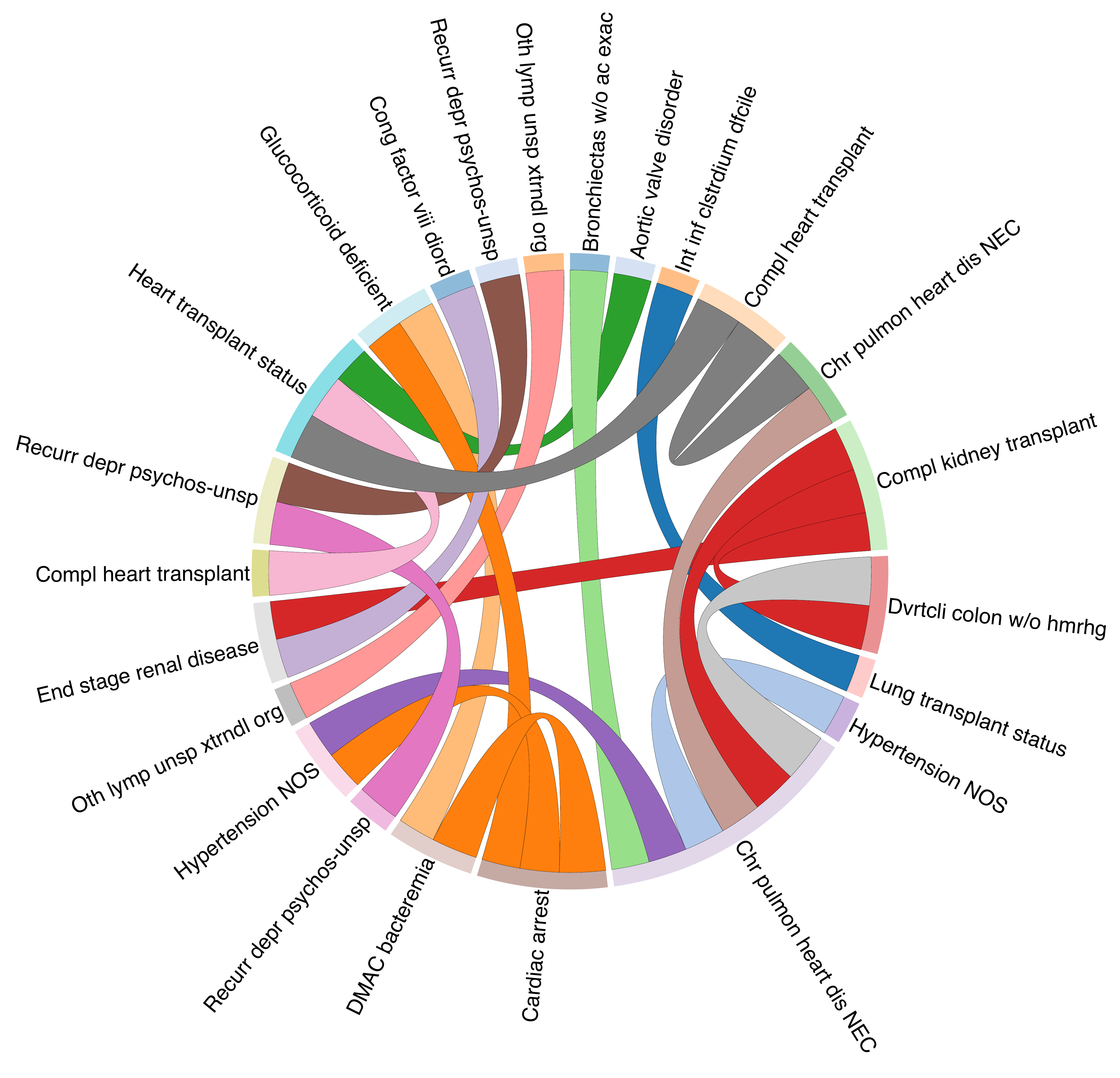}
	\caption{All significant pairwise-positive correlations for `rare' phenotypes (defined as present in less than 5\% of the training set).}
	\label{fig:less_than5}
\end{figure}

In the "more rare" phenotype set, 21 phenotype pairs were found be have a correlation greater than 0.5 (Figure~\ref{fig:less_than5}). These 21 correlations results from 23 unique phenotypes.  Of the 21 reviewed relations 12 (57\%) were found to clinically valid. Most of phenotype pairs that the clinician deemed as unrelated were not very coherent phenotypes which could be expected from phenotypes that were assigned to less of the 5\% of the training set. 

\subsection{Hypothesis 2: focus on HIV phenotypes}

Of the 250 phenotypes, 73 where identified as 'HIV' according to their automatically generated label. The clinical expert evaluation of the these phenotypes showed that most of the identified phenotypes represented a routine primary care visit of an HIV patient. Three phenotypes were clear representations of HIV phenotype and two other phenotypes representing AIDS, the development of HIV into a disease. The rest of the phenotypes were of HIV comorbidities (psychiatric, cancer, renal, neurological, etc) mixed with HIV related observations. A few phenotypes captured behavioral phenotypes (substance abuse) and 11 phenotypes were deemed as non-coherent. 

\subsection{Hypothesis 3: focus on non-HIV phenotypes}
\begin{figure}[h]
	\centering
	\includegraphics[width=\linewidth]{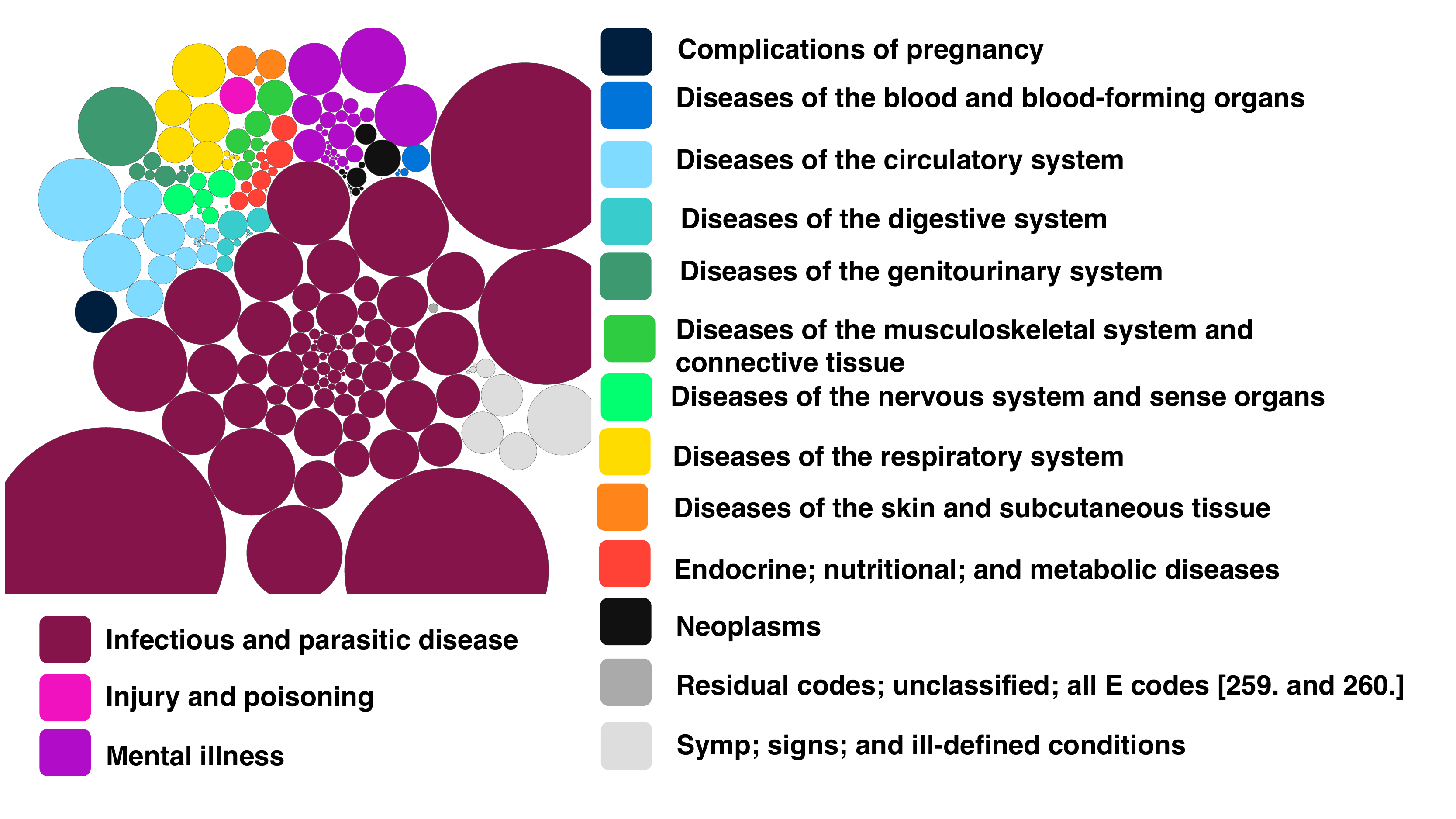}
	\caption{250 learned phenotypes colored by their labels' corresponding CCS category. Size of the circle indicates proportion of phenotype represented in the training set.}
	\label{fig:ccs1}
\end{figure}

When categorized into CCS categories according to their ICD label, the learned phenotypes were were found to cover 16 out of the 18 CCS level 1 classifications (Table~\ref{tab:freq}). The two CCS level 1 categories not captured in the phenotype labels pertained to pediatric conditions. Beyond the most prevalent CCS category related to HIV, `Mental Illness' (which include substance use) and `Disease of the circulatory system' were the most frequent disease groups identified by the model (Figure~\ref{fig:ccs1}). This finding reflects the high coverage of the learned phenotypes related to the types of conditions characteristics of the input population.

\begin{table}
	\caption{250 phenotypes by their CCS category}
	\label{tab:freq}
	\begin{tabular}{lc}
		\toprule
		CCS level 1 Category& Number of \\
		Category& phenotypes\\
		\midrule
		Infectious and parasitic diseases &83
\\
		Mental illness&	32
\\
		Circulatory system&	26\\
		Neoplasms&	23
\\
		Respiratory sys.&	13\\
		Endocrine; nutritional; and metabolic diseases &	12
\\
		Digestive sys.&	10
\\
		Musculoskeletal system and connective tissue&	10\\
		Genitourinary system&	10
\\
		Nervous system and sense organs&	8
\\
		Symp; signs; and ill-defined conditions &	8
\\
		Blood and blood-forming organs&	6
\\
		Injury and poisoning&	4
\\
		Skin and subcutaneous tissue&	3
\\
		Complications of pregnancy &	1
\\
		Resid. codes; unclassified; all E codes&	1
\\
		Certain cond. originating in perinatal period&	0
\\
		Congenital anomalies &	0\\
		\bottomrule
	\end{tabular}
\end{table}

\subsection{Hypothesis 4: types of phenotype- relatedness}
Of the 82 relations evaluated, 63 fit into the same CCS multi-level classification, level 1 category and thus could be inferred using the CCS. However 19 relations were not of the same level 1 category. Out of those 19, 2 were deemed to be unrelated by the clinical expert, 17 relations (21\%) were clinically correct and could not be inferred from the CCS and showed more diversity in the relation type learned:  
the phenotype for severe HIV and one representing the non-disease ICU visits, as well as comorbidity relations like in the pair for `end-stage renal disease' and `acute respiratory failure.'

\subsection{Patient-level summarization}

After the model learns 250 phenotypes from the patient population in the HIV clinic, the model can be applied to the data found in a single patient record. Running the model inference on patient level data (without re-learning the model parameters) provides a 250 dimensional summarization of the patient record. To summarize the patient record over time, we can run the model inference on the patient data after segmenting the patient data at the desired time granularity. The identified phenotype proportions over time is inputed to a sankey visualization presented in Figure~\ref{fig:summary}. Each sankey line represents a phenotype identified to be relevant in the patient record. The hight of each sankey link indicates the proportion of the phenotype in that period. In order to be actionable and avoid information overload the summary showcases the patient's top 5 problems. Top problems are defined at the phenotypes that are found by the model to have the highest probability among the 250 phenotypes learned by the model. The visualization then illustrates how the proportion of the phenotypes increased, decreased, or stayed the same from one period to the next. As a clinical decision support tool, this visualization of the change in phenotypes identified in the patient record could signal to users what health problems the patient possess and how they have changed in salience over time.

The described approach for patient summarizing using the proposed phenotyping model benefits from several of the key characteristics of the model.  Since the phenotyping model is fully unsupervised the model can easily be utilized for other patient populations by re-training the model on relevant patient data. For instance if patient record summarization was desired for oncology patients, the model can be retrained on oncology patients to learn cancer-specific phenotypes as well relevant co-morbidity phenotypes.  The patient summary benefits from the high-throughput nature of the model in that the model learns many phenotypes at the same time and is able to summarize the patient record according to all the phenotypes found to be prevalent in the patient record. Finally the model provides a probabilistic summary of the patient record. The generated patient summary is probabilistic in two senses; 1) each data point has a probability of being associated with the phenotypes; and 2) the phenotype assignments to the patient is also probabilistic which can be interpreted as the salience of the associated phenotype in that time period.

\begin{figure}[h]
	\centering
	\includegraphics[width=\linewidth]{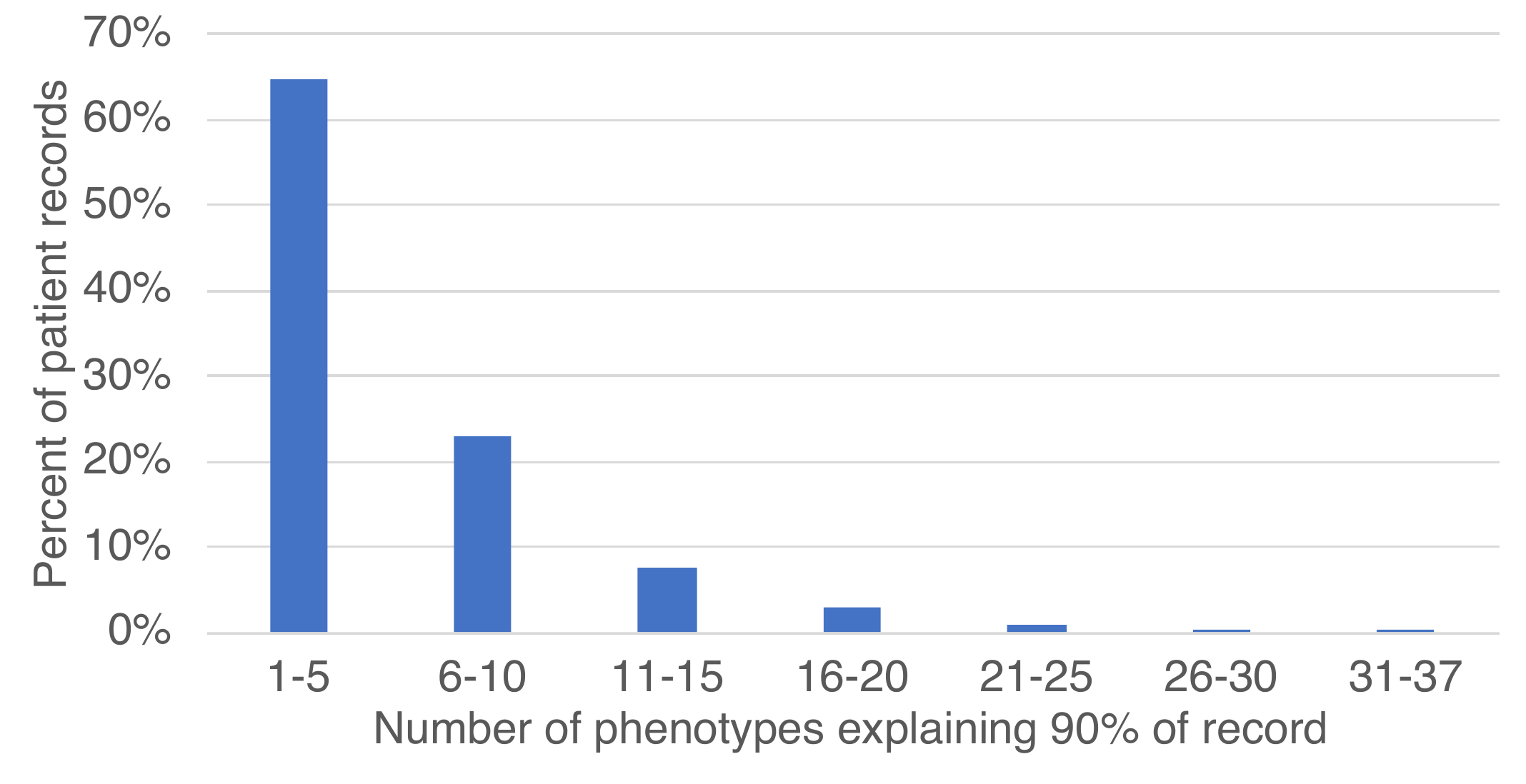}
	\caption{Number of phenotypes needed to explain 90\% of a given patient record. For example, 65\% of the patient records in the training set are almost fully explained (90\% of data) by 1-5 phenotypes. Where each patient record may be explained by a different 1-5 phenotypes from the 250 phenotypes the model learned from the entire patient cohort. }
	\label{fig:ninty}
\end{figure}

To ensure that the model provides a digestible summary of the patient record we analyzed how many phenotypes are required to capture a large majority of the patient record in our training set. Since if we find that the model assigns a large number of phenotypes to each patient record then the proposed summary may still provide too much information be useful at the bedside. In our analysis we found that more than half of patients in our dataset were almost completely described by 1-5 phenotypes (Figure~\ref{fig:ninty}). The large majority of the remaining patients were described by 6-20 phenotypes. Hence, even though is trained to learn a large number of phenotypes (K=250), each patient record is summarized by only a few phenotypes. This indicates that the model has the potential to reduce many thousands of data points in the record of each patient to a list of a handful of problems and how they have changed over time.

\section{Discussion}
Evaluation results show the model simultaneously identifies 250 phenotypes with good coherence and coverage. Learned phenotype relatedness were found clinically meaningful and diverse, identifying some relations out of scope of the baseline resource.  Our experimentation shows that when training the model on a cohort of HIV patients, the model learns multiple HIV phenotypes that can provide good granularity when used for single-patient problem-oriented summarization.  The model was also found to identify a wide range of non-HIV phenotypes, yet commonly encountered in HIV patients. The learned phenotype-phenotype correlations learned from the patient cohort could be used to group and organize highly-related phenotypes in the patient-level summary, to provide a clearer overview of the patient's problems. In many settings but notably urgent care and emergency settings in particular, patient summarization enabled by this model, could provide clinicians a tool for more rapid understanding of the patient comorbidities, leading to better diagnosis, expedited referrals, and potentially a reduction in over-testing. Our future work includes performing an evaluation study of the patient record summarization in assisting clinicians to review patient records more effectively and accurately.

\begin{acks}
   This work was supported by NLM T15 LM007079 Data Science Supplement (GL) and NSF award \#1344668 (NE).
\end{acks}

\bibliographystyle{ACM-Reference-Format}
\bibliography{ctm_bib2}


\section{Online Resources}

Code and phenotype correlations will be made available at publication time. 

\end{document}